\documentclass[aps,twocolumn,prd,preprintnumbers,groupedaddress,nofootinbib,amssymb,eqsecnum,epsfig]{revtex4-1}
\usepackage{graphicx}
\usepackage{bm}
\usepackage{amsmath}
\usepackage{color}
\usepackage{amsfonts}
\usepackage[subnum]{cases}

\begin{document}
\newcommand{\newc}{\newcommand}

\newc{\be}{\begin{equation}}
\newc{\ee}{\end{equation}}
\newc{\ba}{\begin{eqnarray}}
\newc{\ea}{\end{eqnarray}}
\newc{\bea}{\begin{eqnarray*}}
\newc{\eea}{\end{eqnarray*}}
\newc{\D}{\partial}
\newc{\ie}{{\it i.e.} }
\newc{\eg}{{\it e.g.} }
\newc{\etc}{{\it etc.} }
\newc{\etal}{{\it et al.}}
\newcommand{\nn}{\nonumber}
\newc{\ra}{\rightarrow}
\newc{\lra}{\leftrightarrow}
\newc{\lsim}{\buildrel{<}\over{\sim}}
\newc{\gsim}{\buildrel{>}\over{\sim}}
\def\mpl{M_{\rm pl}}
\def\d{\mathrm{d}}

\title{Hairy black-hole solutions in generalized Proca theories}

\author{
Lavinia Heisenberg$^{1}$, 
Ryotaro Kase$^{2}$,
Masato Minamitsuji$^{3}$, and 
Shinji Tsujikawa$^{2}$}

\affiliation{
$^1$Institute for Theoretical Studies, ETH Zurich, Clausiusstrasse 47, 8092 Zurich, Switzerland\\
$^2$Department of Physics, Faculty of Science, Tokyo University of Science, 1-3, Kagurazaka,
Shinjuku-ku, Tokyo 162-8601, Japan\\
$^3$Centro Multidisciplinar de Astrofisica - CENTRA, Departamento 
de Fisica, Instituto Superior Tecnico - IST, Universidade de Lisboa - UL, Avenida 
Rovisco Pais 1, 1049-001 Lisboa, Portugal}

\date{\today}

\begin{abstract}
We present a family of exact black-hole solutions on 
a static spherically symmetric background in second-order 
generalized Proca theories with derivative vector-field 
interactions coupled to gravity. We also derive non-exact
solutions in power-law coupling models including vector 
Galileons and numerically show the existence of 
regular black holes with a primary hair associated 
with the longitudinal propagation. The intrinsic vector-field 
derivative interactions generally give rise to a secondary hair 
induced by non-trivial field profiles. 
The deviation from General Relativity is most significant 
around the horizon and hence there is a golden opportunity for probing 
the Proca hair by the measurements of gravitational waves (GWs) in 
the regime of strong gravity.   
\end{abstract}


\maketitle

\section{Introduction}

The no-hair conjecture of black holes (BHs) \cite{Wheeler} 
was originally suggested by 
the existence of uniqueness theorems for Schwarzschild, 
Reissner-Nordstr\"{o}m (RN), and Kerr solutions 
in General Relativity (GR) \cite{Israel,Carter,Hawking}. 
However, there are several assumptions for proving 
the absence of hairs besides mass, charge, and angular momentum 
in the form of no-hair theorems. 
One of such assumptions for a scalar field $\phi$
is that the standard canonical term $\nabla _{\mu}\phi\nabla^{\mu}\phi/2$ 
is the only field derivative in the action \cite{Bekenstein}. 
Hence, the no-hair theorem of Ref.~\cite{Bekenstein} loses its validity 
for theories containing non-canonical kinetic terms. 

There are theories with non-canonical scalars with non-linear 
derivative interactions-like Galileons \cite{Galileon1,Galileon2} 
and its extension to Horndeski theories \cite{Horndeski,Gao}. 
In shift-symmetric Horndeski theories, a no-hair theorem for static 
and spherically symmetric BHs was proposed \cite{Hui} 
by utilizing the regularity of a Noether current on the horizon.  
A counterexample of a hairy BH evading one of the
conditions discussed in Ref.~\cite{Hui} was advocated for the scalar 
field linearly coupled to a Gauss-Bonnet term \cite{Soti1}. 
For a time-dependent scalar with non-mininal derivative coupling to 
the Einstein tensor, there is also a stealth Schwarzschild solution with a 
non-trivial field profile \cite{Babichev}.

For a massless vector field in GR, the static and spherically 
symmetric BH solution is described by the RN metric with 
mass $M$ and charge $Q$. The introduction of a vector-field 
mass breaks the $U(1)$ gauge symmetry, which allows the 
propagation of the longitudinal mode. 
For this massive Proca field, Bekenstein showed \cite{Bekenstein2} 
that a static BH does not have a vector hair. 
The vector field $A^{\mu}$ vanishes throughout the BH 
exterior from the requirement that a physical scalar 
constructed from $A^{\mu}$ is bounded on a non-singular horizon. 
In this case, the static and spherically symmetric BH solution is described by the Schwarzschild metric 
with mass $M$.

The Bekenstein's no hair theorem \cite{Bekenstein2} cannot be 
applied to the massive vector field 
with non-linear derivative interactions. 
In Refs.~\cite{Heisenberg,Tasinato,Allys,Jimenez2016} 
the action of generalized Proca theories was constructed by demanding 
the condition that the equations of motion are up to
second order to avoid the Ostrogradski instability. 
An exact static and spherically symmetric BH solution with 
the Abelian vector hair\footnote{In this paper we focus on 
the Abelian vector field, but there are hairy 
BH solutions for non-Abelian Yang-Mills fields \cite{colored}. 
A complex Abelian vector field can also give rise to hairy 
Kerr solutions \cite{KerrProca}.}  was found in Ref.~\cite{Chagoya} 
for the Lagrangian 
${\cal L}=(M_{\rm pl}^2/2) R-F_{\mu\nu}F^{\mu\nu}/4
+\beta_4  G^{\mu\nu}A_{\mu}A_{\nu}$ with the specific coupling 
$\beta_4=1/4$, where $M_{\rm pl}$ is the reduced Planck mass,
$R$ and $G_{\mu\nu}$ are the Ricci scalar and Einstein tensor respectively,
and $F_{\mu\nu}=A_{\nu;\mu}-A_{\mu;\nu}$ 
(a semicolon represents a covariant derivative) is the field strength.
This is a stealth Schwarzschild solution containing mass $M$ alone 
with a temporal vector component $A_0=P+Q/r$ 
and a non-vanishing longitudinal component $A_1$, where $r$ is the 
distance from the center of spherical symmetry.
Unlike the RN solution present for the massless vector in GR, 
the Proca hair $P$ is physical but $P$ as well as the electric charge 
$Q$ does not appear in metrics.

The exact BH solutions studied in Ref.~\cite{Chagoya} have been extended 
to non-asymptotically flat solutions \cite{Minami,Babichev17}, 
rotating solutions \cite{Minami}, 
and solutions for $\beta_4\neq 1/4$ \cite{Babichev17,Chagoya2}. 
All these studies considered only the 
$\beta_4 G^{\mu\nu}A_{\mu}A_{\nu}$ coupling. 
It is crucial to investigate whether the self-derivative interactions 
of generalized Proca theories give rise to compelling new hairy BH 
solutions. In this paper we provide a systematic prescription for 
constructing new BH solutions in generalized Proca theories on 
the static and spherically symmetric background given by 
the line element
\be
ds^{2} =-f(r) dt^{2} +h^{-1}(r)dr^{2} + 
r^{2} \left( d\theta^{2}+\sin^{2}\theta\,d\varphi^{2} 
\right)\,,
\label{metric}
\ee
with the vector field $A_\mu=(A_0(r),A_1(r),0,0)$, 
where $f(r)$, $h(r)$, $A_0(r)$, and $A_1(r)$ are arbitrary functions of $r$. 
We will derive exact solutions under the condition 
of a constant norm of the vector field $A_\mu A^\mu={\rm constant}$.
We also numerically obtain hairy BH solutions for power-law coupling 
models including vector Galileons.
Unlike scalar-tensor theories in which hairy BH solutions exist only 
for restrictive cases, we will show that the presence of 
a temporal vector component besides a longitudinal scalar mode 
gives rise to a bunch of hairy BH solutions in broad classes 
of models.

The generalized Proca theories are given by the 
action 
\be
S=\int d^4x \sqrt{-g} \biggl( F+\sum_{i=2}^{6}{\cal L}_i 
\biggr)\,,
\ee
where 
$F=-F_{\mu \nu}F^{\mu \nu}/4$, and \cite{Heisenberg,Jimenez2016}
\ba
& &{\cal L}_2=G_2(X)\,,\qquad 
{\cal L}_3=G_3(X) {A^{\mu}}_{;\mu}\,, \nonumber \\
& &{\cal L}_4=G_4(X)R+G_{4,X} \left[ 
({A^{\mu}}_{;\mu})^2 -{A_{\nu}}_{;\mu} {A^{\mu}}^{;\nu} 
\right]-2g_4(X)F\,, \nonumber \\
& &{\cal L}_5=G_5(X)G_{\mu \nu} {A^{\nu}}^{;\mu}
-\frac{G_{5,X}}{6} [ ({A^{\mu}}_{;\mu})^3
-3{A^{\mu}}_{;\mu} {A_{\sigma}}_{;\rho}
{A^{\rho}}^{;\sigma} \nonumber \\
& &~~~~~+2{A_{\sigma}}_{;\rho}{A^{\rho}}^{;\nu}
{A_{\nu}}^{;\sigma}]
-g_5(X)\tilde{F}^{\alpha \mu} {\tilde{F}^{\beta}}_{\mu} 
A_{\beta;\alpha}\,,\nonumber \\
& &{\cal L}_6=G_6(X) L^{\mu \nu \alpha \beta}
A_{\nu;\mu}A_{\beta;\alpha}
+\frac{G_{6,X}}{2}\tilde{F}^{\alpha \beta} {\tilde{F}^{\mu \nu}} 
A_{\mu;\alpha}A_{\nu;\beta}.
\ea
The functions $G_{2,3,4,5,6}$ and $g_{4,5}$ depend on 
$X=-A_{\mu}A^{\mu}/2$, with the notation 
$G_{i,X}=\partial G_i/\partial X$. The vector field $A^{\mu}$ has 
non-minimal couplings with the Ricci scalar $R$, 
the Einstein tensor $G_{\mu \nu}$, and the  
double dual Riemann tensor 
$L^{\mu \nu \alpha \beta}={\cal E}^{\mu \nu \rho \sigma}
{\cal E}^{\alpha \beta \gamma \delta}R_{\rho \sigma \gamma \delta}/4$, 
where ${\cal E}^{\mu \nu \rho \sigma}$ is the 
Levi-Civita tensor 
and $R_{\rho \sigma \gamma \delta}$ is the Riemann tensor. 
The dual strength tensor $\tilde{F}^{\mu \nu}$ is defined by 
$\tilde{F}^{\mu \nu}={\cal E}^{\mu \nu \alpha \beta}F_{\alpha \beta}/2$.
The Lagrangians containing $g_4,g_5,G_6$ correspond to intrinsic 
vector modes that vanish in the scalar limit 
$A^{\mu} \to \pi^{;\mu}$. 
Throughout the analysis we take into account the 
Einstein-Hilbert term $M_{\rm pl}^2/2$ in $G_4(X)$.

\section{Exact solutions}

The exact solution of Ref.~\cite{Chagoya} was found for 
the model $G_4(X)=M_{\rm pl}^2/2+\beta_4 X$ 
with $\beta_4=1/4$.
For this solution there are two relations
\be
f=h\,,\qquad X=X_c\,,
\label{con}
\ee
where $X_c$ is a constant.
On using these conditions for the vector field 
$A_{\mu}$, it follows that 
\be
A_1=\pm \sqrt{A_0^2-2fX_c}/f\,.
\label{A1}
\ee
Introducing the tortoise coordinate $dr_*=dr/f(r)$, the
scalar product $A_{\mu}dx^{\mu}$ reduces to 
$A_0 du_{\pm}$ around the horizon, where $u_{\pm}=t \pm r_*$. 
The advanced and retarded null coordinates $u_{+}$ 
and $u_{-}$ are regular at the future and past event horizons, 
respectively. Hence the regularity of solutions at the corresponding (future or past) 
horizon is ensured for each branch of 
(\ref{A1}), which is analogous to 
the case of shift-symmetric scalar-tensor theories \cite{Babichev}. 
We will search for exact solutions by 
imposing the two conditions (\ref{con}). 

Provided the condition $G_{4,XX}(X_c)=0$ is satisfied 
for the quartic-order coupling $G_4(X)$, the equation 
of motion for $A_1$ reduces to 
$G_{4,X}(rf'+f-1)A_1=0$, where a prime represents 
the derivative with respect to $r$. 
As long as $G_{4,X} \neq 0$, 
there are two branches characterized by 
$rf'+f-1=0$ or $A_1=0$.
The first gives rise to the stealth Schwarzschild solution 
$f=h=1-2M/r$ found in Ref.~\cite{Chagoya}. 
In this case the temporal vector component obeys 
$A_0''+2A_0'/r=0$, whose integrated solution is 
given by $A_0=P+Q/r$.
Since the constant $P$ is independent of $M$ and $Q$, 
it is regarded as a primary hair \cite{Herdeiro}. 
The other two independent equations are satisfied 
for $G_{4,X}(X_c)=1/4$ and $X_c=P^2/2$, so the longitudinal 
mode (\ref{A1}) reads 
$A_1=\pm \sqrt{2P(MP+Q)r+Q^2}/(r-2M)$.
A concrete model satisfying the above mentioned 
conditions is 
\be
G_4(X)=G_4(X_c)+\frac{1}{4} \left( X-X_c 
\right)+\sum_{n=3} b_n \left( X-X_c \right)^{n}\,,
\label{G4form}
\ee
where $b_n$'s are constants. 
The model $G_4(X)=M_{\rm pl}^2/2+X/4$
corresponds to the special case of Eq.~(\ref{G4form}).

Besides the non-vanishing $A_1$ solution there exists 
another branch $A_1=0$ for the couplings 
$G_i(X)$ with even $i$-index, in which case 
the relation $A_0^2(r)=2f(r)X_c$ 
holds from Eq.~(\ref{A1}).
For the quartic coupling $G_4(X)$
the equation for $A_0$ can be satisfied 
under the conditions $G_{4,X}(X_c)=0$ and 
$2r f f''-rf'^2+4ff'=0$. 
The latter leads to the solution $f=(C-M/r)^2$
with two integrated constants $C$ and $M$. 
For the consistency with the other two equations of motion, 
we require that $C=1$ and $G_{4}(X_c)=X_c/2$. 
Hence we obtain the extremal RN BH solution
\be
f=h=\left( 1-\frac{M}{r} \right)^2,\quad
A_0=P-\frac{PM}{r}\,,\quad A_1=0,
\label{extremum}
\ee
where $P=\pm \sqrt{2X_c}$.
An explicit model realizing this solution is 
\be
G_4(X)=\frac{X_c}{2}
+\sum_{n=2} b_n\left( X-X_c \right)^n\,.
\ee
For the metric (\ref{extremum}), 
$P$ depends on $M$ by reflecting the fact that 
the charge $Q=-PM$ has a special relation with 
the mass $M$. Hence the Proca hair is of the 
secondary type.

For the cubic coupling $G_3(X)$ the equation 
for $A_1$ reads 
\be
G_{3,X} \left[ f^2 (rf'+4f)A_1^2
+rA_0 (2fA_0'-f'A_0) \right]=0\,,
\label{G3Xeq}
\ee
so there are two branches satisfying  
(i) $G_{3,X}(X_c)=0$ or 
(ii) $G_{3,X}(X_c) \neq 0$.
For the branch (i) the consistency with the other equations 
requires that $2 \left( rf'+f-1 \right)M_{\rm pl}^2
+r^2 A_0'^2=0$ and $A_0''+2A_0'/r=0$, so 
the integrated solutions are of the RN forms:
\be
f=h=1-\frac{2M}{r}+\frac{Q^2}{2M_{\rm pl}^2 r^2}\,,\quad
A_0=P+\frac{Q}{r}\,,
\label{G3fhA0}
\ee
with the non-vanishing longitudinal mode (\ref{A1}).
This exact solution can be realized by the model 
\be
G_3(X)=G_3(X_c)+\sum_{n=2} b_n\left( X-X_c \right)^n\,.
\label{G3exactlag}
\ee
Unlike the RN solution in GR with $G_3(X)=0$, 
$P$ in Eq.~(\ref{G3fhA0}) has the meaning 
of the primary hair with the non-vanishing longitudinal 
mode (\ref{A1}).
The branch (ii) corresponds to the case in which the terms 
in the square bracket of Eq.~(\ref{G3Xeq}) vanishes. 
On using Eq.~(\ref{A1}) and imposing the asymptotically 
flat boundary condition $f \to 1$ for $r \to \infty$, 
we obtain the extremal RN BH solution (\ref{extremum}) with 
$P=\pm \sqrt{2}M_{\rm pl}$.

For the quintic coupling $G_5(X)$ the temporal component 
obeys $A_0''+2A_0'/r=0$ under the conditions (\ref{con}), 
so the resulting solution is $A_0=P+Q/r$. 
Imposing the condition $G_{5,X}(X_c)=0$ further, 
the equation for $A_1$ reduces to 
$(A_0A_0'-X_cf')A_1^2G_{5,XX}=0$ and hence 
there are two branches satisfying 
(i) $A_0A_0'=X_c f'$ or (ii) $A_1=0$. 
For the branch (i),  the resulting solutions are given by 
the RN solutions (\ref{G3fhA0}) with the particular 
relations $P=-2MM_{\rm pl}^2/Q$ and $X_c=M_{\rm pl}^2$.
The longitudinal mode (\ref{A1}) reduces to
\be
A_1=\pm \frac{2M_{\rm pl}^3 
\sqrt{2(2M^2M_{\rm pl}^2-Q^2)}\,r^2}
{Q[2M_{\rm pl}^2r(2M-r)-Q^2]}\,,
\label{A1G5}
\ee
whose existence requires the condition 
$2M^2M_{\rm pl}^2>Q^2$. 
Since $P$ depends on $M$ and $Q$, the 
Proca hair $P$ is secondary. 
This exact solution can be realized by the model 
\be
G_5(X)=G_5(X_c)+\sum_{n=2} b_n 
\left( X-M_{\rm pl}^2 \right)^n\,.
\label{G5explicit}
\ee
Another branch $A_1=0$ is the special 
case of Eq.~(\ref{A1G5}), i.e., $Q^2=2M^2M_{\rm pl}^2$, 
under which the solution is given by the extremal 
RN BH solution (\ref{extremum}) 
with $P=\pm \sqrt{2}M_{\rm pl}$.

The sixth-order coupling $G_6(X)$ has the two branches 
(i) $A_1=0$ or (ii) $A_0'=0$. 
For the branch (i) there exists an exact solution if the two 
conditions $G_6(X_c)=0$ and $G_{6,X}(X_c)=0$ hold.
This is the extremal RN BH solution (\ref{extremum})
with $X_c=M_{\rm pl}^2$ and $P=\pm \sqrt{2}M_{\rm pl}$, 
which can be realized for the model
\be
G_6(X)=\sum_{n=2} b_n \left( X-M_{\rm pl}^2 \right)^n\,.
\ee
The branch (ii) corresponds to $A_0={\rm constant}$, 
in which case we obtain the stealth Schwarzschild solution 
$f=h=1-2M/r$. This exists for general couplings $G_6(X)$ 
with arbitrary values of $A_1$. Since we are now imposing  
the second condition of Eq.~(\ref{con}), the longitudinal mode is fixed to be 
$A_1=\pm \sqrt{r[(A_0^2-2X_c)r+4MX_c]}/(r-2M)$.

\section{Power-law couplings}

So far we have imposed the conditions 
(\ref{con}) to derive exact solutions, but we will 
also study BH solutions for the power-law models
\be
G_i(X)=\tilde{\beta}_i X^n\,,\qquad 
g_j(X)=\tilde{\gamma}_{j}X^n\,,
\label{powerlaw}
\ee
where $n$ is a positive integer, and $\tilde{\beta_i}$ and 
$\tilde{\gamma}_j$ are coupling constants\footnote{
For the dimensionless coupling constants, we use the 
notations $\beta_i$ and 
$\gamma_j$ in the following.} 
with $i=3,4,5,6$ and $j=4,5$.

Let us begin with the cubic vector-Galileon 
interaction $G_3(X)=\beta_3 X$.
Then, the longitudinal mode obeys
\be
A_1=\pm \sqrt{\frac{rA_0(f'A_0-2fA_0')}
{fh(rf'+4f)}}\,.
\label{A1G3}
\ee
Around the horizon characterized by the radius $r_h$, 
we expand $f,h,A_0$ in the forms 
\ba
& &
f=\sum_{i=1}^{\infty} f_i(r-r_h)^i\,,\qquad
h=\sum_{i=1}^{\infty} h_i(r-r_h)^i\,,
\nonumber \\ 
& &
A_0=a_0+\sum_{i=1}^{\infty} a_i(r-r_h)^i\,,
\label{fhorizon}
\ea
where $f_i,h_i,a_0,a_i$ are constants. 
To recover the RN solutions of the form 
$f=h=(r-r_h)(r-\mu r_h)/r^2$ in the limit  
$\beta_3 \to 0$, 
where the constant $\mu$ is in the range $0<\mu<1$
so that $r=r_h$ corresponds to the outer horizon, 
we choose
\be
f_1=h_1=(1-\mu)/r_h\,.
\label{f1h1}
\ee
Taking the positive branch of $A_1$ with $a_0>0$ 
and picking up linear-order terms in $\beta_3$, 
the effect of the coupling $\beta_3$ starts to appear 
at second order of $(r-r_h)^i$, such that 
\ba
& & 
a_1=\frac{M_{\rm pl}\sqrt{2\mu}}{r_h}\,,\qquad
a_2=-\frac{M_{\rm pl}\sqrt{2\mu}}{r_h^2}
+\alpha_2 \beta_3\,,\nonumber \\
& &
f_2=\frac{2\mu-1}{r_h^2}+{\cal F}_2 \beta_3\,,
\quad~
h_2=\frac{2\mu-1}{r_h^2}+{\cal H}_2 \beta_3\,,
\label{f2h2}
\ea
where $\alpha_2, {\cal F}_2, {\cal H}_2$ depend on 
the three parameters $(h_1, r_h, a_0)$.
The coupling $\beta_3$ induces the difference between 
the metrics $f$ and $h$.
The leading-order longitudinal mode around $r=r_h$
is given by $A_1=a_0/[f_1(r-r_h)]$, 
so the scalar product $A_{\mu}dx^{\mu}$ 
becomes
$A_{\mu}dx^{\mu} \simeq a_0du_+$, 
which is regular at the future horizon $r=r_h$. 

We also search for asymptotic flat solutions at spatial 
infinity ($r \to \infty$) by expanding $f,h,A_0$ in the forms
\be
f=1+\sum_{i=1}^{\infty}\frac{\tilde{f}_i}{r^i},\quad
h=1+\sum_{i=1}^{\infty}\frac{\tilde{h}_i}{r^i},\quad
A_0=P+\sum_{i=1}^{\infty}\frac{\tilde{a}_i}{r^i}\,.
\label{fh}
\ee
For the cubic Galileon, the asymptotic solution for $A_1$ 
reduces to $A_1=\sum_{i=2}^{\infty} \tilde{b}_i/r^i$, 
where the first-order coefficient $\tilde{b}_1$ 
vanishes from the background equations of motion.
The iterative solutions are given by 
$f=1-2M/r-P^2M^3/(6M_{\rm pl}^2r^3)+{\cal O}(1/r^4)$, 
$h=1-2M/r-P^2M^2/(2M_{\rm pl}^2r^2)
-P^2M^3/(2M_{\rm pl}^2r^3)+{\cal O}(1/r^4)$, and 
$A_0=P-PM/r-PM^2/(2r^2)+{\cal O}(1/r^3)$, where 
we have set $\tilde{f}_1=\tilde{h}_1=-2M$.
The coefficient $ \tilde{b}_2$ and the 
coupling $\beta_3$ begin to appear at the orders of 
$1/r^4$ and $1/r^5$, respectively, in $f,h,A_0$.

\begin{figure}
\begin{center}
\includegraphics[height=3.3in,width=3.2in]{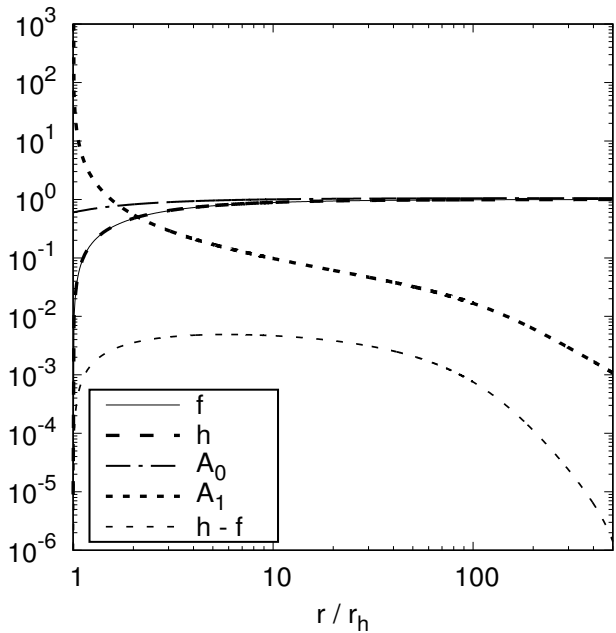}
\end{center}
\caption{\label{fig1}
Numerical solutions of $f, h,A_0, A_1,h-f$ outside the horizon 
for the cubic vector-Galileon model $G_3(X)=\beta_3X$ 
with $\beta_3=10^{-3}/(r_h M_{\rm pl})$.
The boundary conditions around the horizon are chosen to satisfy Eqs.~(\ref{fhorizon})-(\ref{f2h2}) with $\mu=0.1$ 
and $a_0=0.6M_{\rm pl}$ at $r=1.001r_h$.}
\end{figure}

In Fig.~\ref{fig1} we plot one example of numerically 
integrated solutions outside the horizon derived by using 
the boundary conditions (\ref{fhorizon})-(\ref{f2h2}) 
around $r=r_h$. The solutions in the two asymptotic 
regimes smoothly join each other without any 
discontinuity. As estimated above, the longitudinal mode 
behaves as $A_1 \propto (r-r_h)^{-1}$ for $r \simeq r_h$ 
and $A_1 \propto r^{-2}$ for $r \gg r_h$. 
Since the time $t$ can be reparametrized such that   
$f$ shifts to $1$ at spatial infinity, we have 
performed this rescaling of $f$ after solving the equations 
of motion up to $r=2 \times 10^{7}r_h$. In Fig.~\ref{fig1} the 
difference between $h$ and $f$ manifests itself in the 
regime of strong gravity with the radius $r \lesssim 100 r_h$.

Since the two asymptotic solutions discussed above are continuous, 
the three parameters $(\tilde{b}_2, M, P)$ 
appearing in the expansion (\ref{fh}) with 
 $A_1=\sum_{i=2}^{\infty} \tilde{b}_i/r^i$ are related to 
the three parameters $(h_1, r_h, a_0)$ arising 
in the expansion (\ref{fhorizon}), as 
$\tilde{b}_2=\tilde{b}_2(h_1, r_h, a_0)$,
$M=M(h_1, r_h, a_0)$, and 
$P=P(h_1,r_h,a_0)$.
Since $\tilde{b}_2$ is not fixed by the two parameters 
$M$ and $P$ alone, this is regarded as a primary hair.

For the cubic interaction $G_3(X)=\beta_3 M_{\rm pl}^2 
(X/M_{\rm pl}^2)^n$ with $n \geq 2$, there is the 
non-vanishing $A_1$ branch 
satisfying the relation (\ref{A1G3}).
In this case, the property of two asymptotic solutions (\ref{fhorizon}) and (\ref{fh}) is similar to that discussed 
for $n=1$. The solutions are also regular throughout the 
BH exterior with the difference between $f$ and 
$h$ induced by $\beta_3$. 
There is also another branch obeying 
\be
A_1=\pm \sqrt{A_0^2/(fh)}\,,
\label{A1RN}
\ee
for which the resulting solutions correspond to the RN 
solutions (\ref{G3fhA0}). Indeed, this exact solution 
is the special case of the model (\ref{G3exactlag}) with 
$X_c=0$ and $G_3(X_c)=0$.

Let us proceed to the quartic coupling 
$G_4(X)=\beta_4 M_{\rm pl}^2 (X/M_{\rm pl}^2)^n$ 
with $n \geq 2$. In general, we have two branches 
characterized by (i) $A_1 \neq 0$ or (ii) $A_1=0$.
For $n \geq 3$ there exists the non-vanishing $A_1$ 
branch (\ref{A1RN}) with the RN solutions (\ref{G3fhA0}).
Another non-vanishing $A_1$ branch gives rise to 
hairy BH solutions with $f \neq h$. 
Indeed, the solutions around the horizon are given by 
the expansion (\ref{fhorizon}) with the coupling 
$\beta_4$ appearing at second order ($i=2$) in $f, h$ 
and at first order ($i=1$) in $A_0$. 
They are characterized by the three parameters 
$(h_1, r_h,a_0)$ under the condition (\ref{f1h1}).
The solutions expanded at spatial infinity are 
the RN solutions (\ref{G3fhA0}) with corrections 
induced by $\beta_4$. 
If $n=2$, for example, such corrections to $f,h,A_0$ 
arise at second order in $1/r^2$, e.g., 
$\delta f=3P^2Q^2(5P^2-8M_{\rm pl}^2)\beta_4/
(4M_{\rm pl}^6 r^2)$, 
$\delta h=3P^2Q^2(11P^2-16M_{\rm pl}^2)\beta_4/
(4M_{\rm pl}^6 r^2)$, and 
$\delta A_0=PQ^2 (3P^2-4M_{\rm pl}^2)\beta_4/
(M_{\rm pl}^4 r^2)$, respectively.
The longitudinal mode behaves as 
$A_1 \propto (r-r_h)^{-1}$ for $r \simeq r_h$ and 
$A_1 \propto r^{-1/2}$ for $r \gg r_h$. 
Numerically we confirmed that the solution around 
$r=r_h$ smoothly connects to that in the 
asymptotic regime $r \gg r_h$, so the parameters 
$(P,Q,M)$ are related to $(h_1, r_h,a_0)$. 
Therefore, the Proca hair $P$ is of the primary type.

For the second branch $A_1=0$, the solutions 
(\ref{fhorizon}) around $r=r_h$ are subject to 
the constraint $a_0=0$. Hence they are expressed in 
terms of the two parameters $(h_1,r_h)$
with the coupling $\beta_4$ appearing at the order 
of $(r-r_h)^3$ in $f,h,A_0$ for $n=2$.
At spatial infinity, the effect of $\beta_4$ works as 
corrections to the RN solutions (\ref{G3fhA0}). 
For $n=2$ the leading-order corrections to $f,h,A_0$ 
are given, respectively, by 
$\delta f=-4P^3(MP+Q)\beta_4/(M_{\rm pl}^4r)$, 
$\delta h=3P^4Q^2\beta_4/(4M_{\rm pl}^6 r^2)$, and 
$\delta A_0=-P^3Q(2MP+Q)\beta_4/(2M_{\rm pl}^4r^2)$. 
The matching of two asymptotic solutions has been also 
confirmed numerically, so $(P,M,Q)$ depend on the 
two parameters $(h_1, r_h)$ alone. Hence 
$P$ corresponds to the secondary hair.

The sixth-order coupling 
$G_6(X)=(\beta_6/M_{\rm pl}^2 )(X/M_{\rm pl}^2)^n$ 
with the power $n \geq 0$ has the branch satisfying 
$A_1^2/A_0^2=(3h-1)/[fh\{(2n+1)h-1\}]$ 
besides $A_0'=0$, $A_1=0$, and 
$A_1=\pm \sqrt{A_0^2/(fh)}$ (the last one is present for $n \geq 3$). 
However, the first one does not  
exist in the region $1/(2n+1)<h<1/3$ outside the horizon.
Since the second and fourth branches correspond to the 
Schwarzschild and RN solutions, respectively, 
the branch $A_1=0$ alone leads to the solutions 
with $f \neq h$. 
The $U(1)$-invariant interaction derived by 
Horndeski 
\cite{Horndeski76} corresponds to $n=0$, 
in which case the coupling $\beta_6$ appears in the 
expansion (\ref{fhorizon}) around $r=r_h$ at second order 
for $f,h$ and at first order for $A_0$, with $a_0$ unfixed. 
For $n \geq 1$ the effect of $\beta_6$ arises at $n+1$ 
order in Eq.~(\ref{fhorizon}), with $a_0=0$.
At spatial infinity, the leading-order corrections to the 
RN solutions (\ref{G3fhA0}) read
$\delta f=-P^{2n}Q^2\beta_6/(2^{1+n}M_{\rm pl}^{4+2n}r^4)$, 
$\delta h=(2n-1)MP^{2n}Q^2\beta_6/(2^{1+n}M_{\rm pl}^{4+2n}r^5)$, 
and
 $\delta A_0=-MP^{2n}Q \beta_6/(2^nM_{\rm pl}^{2+2n}r^4)$,
which match with those
derived by Horndeski in the $U(1)$-invariant case ($n=0$) \cite{HorndeskiBH}. 
For $n \geq 0$, the numerically integrated solutions are regular 
throughout the horizon exterior with the difference between 
$f$ and $h$. When $n=0$, 
$P$ has no physical meaning 
due to the $U(1)$ gauge symmetry, so there are two physical hairs $M$ and $Q$ 
related to the parameters $h_1$ and $r_h$ around the horizon. 
For $n \geq 1$ the Proca hair $P$ is secondary, which
reflects the fact that $(P,M,Q)$ depend on $(h_1,r_h)$ alone.

The quintic coupling $G_5(X)=\beta_5 (X/M_{\rm pl}^2)^n$ does not 
lead to regular solutions with $A_1 \neq 0$ due to the divergence 
at $h=1/(2n+1)$. For the intrinsic vector-mode couplings 
$g_4(X)=\gamma_4(X/M_{\rm pl}^2)^n$ and 
$g_5(X)=(\gamma_5/M_{\rm pl}^2)(X/M_{\rm pl}^2)^n$ 
with $n \geq 1$, there are hairy regular BH solutions 
with $f \neq h$ characterized by $A_1=0$ and 
$A_1=\pm \sqrt{A_0^2/[(1+2n)fh]}$, 
respectively. The couplings $\gamma_4$ 
and $\gamma_5$ 
give rise to corrections to the RN solutions (\ref{G3fhA0}), 
where the near-horizon expansion (\ref{fhorizon}) can be 
expressed in terms of the two parameters $h_1$ and $r_h$
with $a_0=0$.
In this case the Proca hair $P$ is dependent on $M$ and 
$Q$ at spatial infinity, so it is of the secondary type.

\section{Conclusions}

We have systematically constructed new exact BH solutions 
under the conditions (\ref{con}) and also obtained a family of hairy 
numerical BH solutions with $f \neq h$ for the power-law models (\ref{powerlaw}). 
For the cubic and quartic couplings 
$G_3(X)=\tilde{\beta}_3 X^n$ and $G_4(X)=\tilde{\beta}_4 X^n$, 
there exist non-vanishing $A_1$ branches with the primary Proca hair 
with the difference between $f$ and $h$ manifesting around the horizon. 
For the intrinsic vector-mode couplings 
$G_6(X)=\tilde{\beta}_6 X^n$, $g_4(X)=\tilde{\gamma}_4X^n$, 
$g_5(X)=\tilde{\gamma}_5X^n$ with $n \geq 1$, 
there are regular BH solutions (RN solutions 
with corrections induced by the couplings)
characterized by the secondary 
Proca hair $P$.

Since astronomical observations of BHs have increased their accuracies, 
there will be exciting possibilities for probing deviations from GR in the 
foreseeable future, e.g., in the measurements of innermost stable circular orbits. 
GWs emitted from quasi-circular BH binaries 
can generally place tight bounds on modified gravitational theories with large
deviations from GR in the regime of strong gravity \cite{Yagi}. 
The future GW measurements 
will be able to measure 
the Proca charge $P$ through the corrections to the Schwarzschild or 
RN metrics
and the precise determination of polarizations.
The existence of such a new vector hair will shed new light 
on the construction of unified theories connecting gravitational theories 
with particle theories.
Our analysis in the strong gravity regime is also
complementary to the cosmological analysis with the late-time 
acceleration \cite{cosmo} and the solar-system constraints \cite{screening}. 
The combination of them will allow us to probe vector-tensor theories
in all scales in astrophysics and cosmology. 

\section*{ACKNOWLEDGMENTS}
L.\,H. thanks financial support from Dr.~Max R\"ossler, 
the Walter Haefner Foundation and the ETH Zurich
Foundation. R.\,K. is supported by the Grant-in-Aid for Young Scientists B of the JSPS No.\,17K14297. 
M.\,M. is supported by FCT-Portugal 
through Grant No. SFRH/BPD/88299/2012. 
S.\,T. is supported by the Grant-in-Aid for Scientific Research Fund of the JSPS No.~16K05359 and 
MEXT KAKENHI Grant-in-Aid for 
Scientific Research on Innovative Areas ``Cosmic Acceleration'' (No.\,15H05890).



\begin{thebibliography}{99}

\bibitem{Wheeler} 
R.~Ruffini and J.~A.~Wheeler, Phys.\ Today {\bf 24}, 
No. 1, 30 (1971).

\bibitem{Israel} 
W.~Israel,
Phys.\ Rev.\  {\bf 164}, 1776 (1967).

\bibitem{Carter} 
B.~Carter,
Phys.\ Rev.\ Lett.\  {\bf 26}, 331 (1971).

\bibitem{Hawking} 
S.~W.~Hawking,
Commun.\ Math.\ Phys.\  {\bf 25}, 152 (1972).

\bibitem{Bekenstein} 
J.~D.~Bekenstein,
Phys.\ Rev.\ D {\bf 51}, R6608 (1995).

\bibitem{Galileon1} 
A.~Nicolis, R.~Rattazzi and E.~Trincherini,
Phys.\ Rev.\ D {\bf 79}, 064036 (2009).

\bibitem{Galileon2} 
C.~Deffayet, G.~Esposito-Farese and A.~Vikman,
Phys.\ Rev.\ D {\bf 79}, 084003 (2009).

\bibitem{Horndeski} 
G.~W.~Horndeski,
Int.\ J.\ Theor.\ Phys.\  {\bf 10}, 363 (1974).

\bibitem{Gao} 
C.~Deffayet, X.~Gao, D.~A.~Steer and G.~Zahariade,
Phys.\ Rev.\ D {\bf 84}, 064039 (2011).

\bibitem{Hui} 
L.~Hui and A.~Nicolis,
Phys.\ Rev.\ Lett.\  {\bf 110}, 241104 (2013).

\bibitem{Soti1} 
T.~P.~Sotiriou and S.~Y.~Zhou,
Phys.\ Rev.\ Lett.\  {\bf 112}, 251102 (2014).

\bibitem{Babichev} 
E.~Babichev and C.~Charmousis,
JHEP {\bf 1408}, 106 (2014).

\bibitem{Bekenstein2} 
J.~D.~Bekenstein,
Phys.\ Rev.\ D {\bf 5}, 1239 (1972).

\bibitem{Heisenberg} 
L.~Heisenberg,
JCAP {\bf 1405}, 015 (2014).
 
\bibitem{Tasinato}
G.~Tasinato,
JHEP {\bf 1404}, 067 (2014).
 
\bibitem{Allys}
E.~Allys, P.~Peter and Y.~Rodriguez,
JCAP {\bf 1602}, 004 (2016).
 
\bibitem{Jimenez2016} 
J.~B.~Jimenez and L.~Heisenberg,
Phys.\ Lett.\ B {\bf 757}, 405 (2016).

\bibitem{colored}
R.~Bartnik and J.~Mckinnon,
Phys.\ Rev.\ Lett.\  {\bf 61}, 141 (1988);
M.~S.~Volkov and D.~V.~Galtsov,
JETP Lett.\  {\bf 50}, 346 (1989);
P.~Bizon,
Phys.\ Rev.\ Lett.\  {\bf 64}, 2844 (1990).

\bibitem{KerrProca} 
C.~Herdeiro, E.~Radu and H.~Runarsson,
Class.\ Quant.\ Grav.\  {\bf 33}, 154001 (2016).

\bibitem{Chagoya} 
J.~Chagoya, G.~Niz and G.~Tasinato,
Class.\ Quant.\ Grav.\  {\bf 33}, 175007 (2016).

\bibitem{Minami} 
M.~Minamitsuji,
Phys.\ Rev.\ D {\bf 94}, 084039 (2016).

\bibitem{Babichev17} 
E.~Babichev, C.~Charmousis and M.~Hassaine,
JHEP {\bf 1705}, 114 (2017).

\bibitem{Chagoya2} 
J.~Chagoya, G.~Niz and G.~Tasinato,
Class.\ Quant.\ Grav.\  {\bf 34}, no. 16, 165002 (2017).

\bibitem{Herdeiro} 
C.~A.~R.~Herdeiro and E.~Radu,
Int.\ J.\ Mod.\ Phys.\ D {\bf 24}, 1542014 (2015).

\bibitem{Horndeski76} 
G.~W.~Horndeski,
J.\ Math.\ Phys.\  {\bf 17}, 1980 (1976).

\bibitem{HorndeskiBH} 
G.~W.~Horndeski,
Phys.\ Rev.\ D {\bf 17}, 391 (1978).

\bibitem{Yagi} 
K.~Yagi, N.~Yunes and T.~Tanaka,
Phys.\ Rev.\ Lett.\  {\bf 109}, 251105 (2012).

\bibitem{cosmo} 
A.~De Felice, L.~Heisenberg, R.~Kase, S.~Mukohyama, 
S.~Tsujikawa and Y.~l.~Zhang,
JCAP {\bf 1606}, 048 (2016).

\bibitem{screening} 
A.~De Felice, L.~Heisenberg, R.~Kase, S.~Tsujikawa, 
Y.~l.~Zhang and G.~B.~Zhao,
Phys.\ Rev.\ D {\bf 93}, 104016 (2016).

\end{thebibliography}
\end{document}